\documentclass[3p,times,twocolumn]{elsarticle}

\usepackage{ecrc}

\volume{00}

\firstpage{1}

\journalname{Nuclear Physics B Proceedings Supplement}

\runauth{}

\jid{nppp}

\jnltitlelogo{Nuclear Physics B Proceedings Supplement}

\usepackage{graphicx}
\usepackage{bm,amsmath,amssymb}
\usepackage[figuresright]{rotating}

\newcommand{\beq}{\begin{equation}}
\newcommand{\eeq}{\end{equation}}

\newcommand{\bp}{\bm{p}}
\newcommand{\bx}{\bm{x}}

\newcommand{\mcal}{\mathcal}
\newcommand{\rmd}{{\rm d}}
\newcommand{\rme}{{\rm e}}
\newcommand{\eqn}[1]{Eq.~(\ref{#1})}

\newcommand{\tth}{t_{\rm drag}}
\newcommand{\tbr}{t_{\rm br}}
\newcommand{\trel}{t_{\rm rel}}
\newcommand{\tf}{t_{\rm f}}

\begin{document}

\begin{frontmatter}

\title{Medium-induced jet evolution: multiple branching and thermalization}
   \author{Edmond Iancu\corref{cor1}}
   \ead{edmond.iancu@cea.fr}
   
   \author{Bin Wu}
\ead{bin.wu@cea.fr}

  \cortext[cor1]{Corresponding author}
 \address{Institut de Physique Th\'{e}orique, CEA Saclay, CNRS UMR 3681, F-91191 Gif-sur-Yvette, France}


\title{Medium-induced jet evolution: multiple branching and thermalization}

\begin{abstract}
For an energetic jet propagating through a weakly-coupled quark-gluon plasma,
we present the physical picture of jet quenching in longitudinal phase-space, as 
emerging from the interplay between the medium-induced, quasi-democratic,
branchings and the elastic collisions responsible for the thermalization of the soft
branching products.

\end{abstract}


\end{frontmatter}

\section{Introduction}
The experimental studies of jets in Pb+Pb collisions at the LHC have triggered intense theoretical 
efforts towards understanding global observables 
like the shape of a jet propagating through a dense QCD medium 
and the energy transfer by the jet towards the medium.
These efforts lead to the emergence of a new picture for the in-medium jet evolution, 
in which the energy of the hard components of the jet is efficiently transmitted, 
via multiple, quasi-democratic, branchings,
to a multitude of comparatively soft gluons, which are then deviated 
towards large angles by rescattering in the medium
\cite{Blaizot:2012fh,Blaizot:2013hx,Fister:2014zxa,Kurkela:2014tla,Apolinario:2014csa,Iancu:2015uja,Blaizot:2015lma}. In particular, Ref.~\cite{Iancu:2015uja} presented for the first time the picture of this
evolution in longitudinal phase-space, with the longitudinal axis defined as the
direction of propagation of the leading particle. This picture
will be succinctly summarized in what follows.
\label{sec:Intro}

\section{Kinetic theory for jet evolution}
\label{sec:kin}

We study the parton distribution produced by a high-energy jet propagating
through a weakly-coupled quark-gluon plasma in thermal equilibrium at temperature $T$. 
The leading particle (LP) which initiates the jet has a high
energy $E\gg T$ and a comparatively small virtuality. 
We concentrate on the {\em medium-induced} evolution, as
triggered by the collisions between the partons from the jet and the constituents of the medium.
One can distinguish between two types of collisions:

\texttt{(i)} elastic, $2\! \to\! 2$, collisions, which entail energy and momentum transfer between
the jet and the medium;

\texttt{(ii)} inelastic collisions, like $2\to 3$ or, more generally, 1+(many)$\to$ 2+(many), in which
a parton from the jet undergoes a $1\to 2$ branching.

At weak coupling, such processes can be described by a kinetic equation for the gluon
distribution \cite{Baier:2000sb,Arnold:2002zm},
\beq\label{generaleq}
\left(\frac{\partial}{\partial t}+{\bm v}\cdot \nabla_{\bm x} \right)f(t,\bx,\bp)\, =\, \mcal{C}_{\rm el}[f]+
\mcal{C}_{\rm br}[f]\,.\eeq
Here, $f(t,\bx,\bp)$ is the gluon occupation number, 
${\bm v}=\bp/p$ with $p\equiv |\bp|$ is the gluon velocity, and $\mcal{C}_{\rm el}[f]$ 
and $\mcal{C}_{\rm br}[f]$ are collision terms which encompass the elastic and inelastic processes,
respectively. Explicit expressions can be found in \cite{Arnold:2002zm,Huang:2014iwa,Ghiglieri:2015zma},
but the general equation is too complicated to be solved in practice, even numerically. In what
follows, we shall examine the relevant processes in mode detail, in order
to motive a simple approximation
to \eqn{generaleq} which is tractable in practice \cite{Iancu:2015uja}. 

The medium-induced parton cascades are controlled by relatively
hard gluons, with momenta $p$ within the range $T\ll p \ll E$, which 
undergo small-angle  scattering. 
The elastic collision term can therefore be evaluated in the Fokker-Planck approximation
\cite{Huang:2014iwa,Ghiglieri:2015zma}:
\beq\label{CFP}
\mcal{C}_{\rm el}[f]\,\simeq\,
\frac{1}{4}\,\hat{q}\,\nabla_{\bp}\cdot\left[ \left( \nabla_{\bp} + \frac{{\bm v}}{T}\right) f\right]\,,
\eeq
with $\hat q\!\sim\! \alpha_s^2 T^3\ln(1/\alpha_s)$ 
the jet quenching parameter. 

Consider a test particle which at $t=0$ has a high momentum $p_0\gg T$ oriented along the
$z$-axis.  \eqn{CFP} implies that this particle loses energy according to
$\langle p_z(t)\rangle\simeq p_0 -\eta t$, with $\eta\equiv \hat q/4T$  (the `drag coefficient'),
and at the same time suffers transverse and longitudinal momentum broadening:
$\langle p^2_\perp\rangle\simeq \hat q t$ and $\langle \Delta p^2_z\rangle\simeq (\hat q/2) t$. 
This dynamics eventually drives the test
particles to thermal equilibrium. Indeed, one can easily check that \eqn{CFP} admits the
Maxwell-Boltzmann thermal distribution $f_{\rm eq}\propto \rme^{-p/T}$  as a fixed point.

In the absence of inelastic process, the test particle would lose its initial energy $p_0$
via drag after a time $\tth(p_0)\simeq p_0/\eta$ and then thermalize via
diffusive processes under an additional time $\Delta t\sim \trel$, with
\beq\label{trel}
\trel\equiv \frac{4T^2}{\hat{q}}\,\sim\, \frac{1}{\alpha_s^2 T\ln(1/\alpha_s)}\,.
\eeq
Observing that $\tth(p_0)\simeq (p_0/T)\trel \gg \trel$ when $p_0\gg T$, it becomes
clear that the overall duration of the thermalization process
is controlled by the first stage --- the energy loss via drag.

But after adding the inelastic collisions, i.e. the term $\mcal{C}_{\rm br}[f]$
in the r.h.s. of  \eqref{generaleq},
the above scenario changes dramatically, at least for
the relatively hard components of the jet with $p\gg T$. 
This is so since multiple branching is much more efficient 
than elastic collisions in redistributing the energy among the soft modes. 

Indeed, in the presence of inelastic collisions, a gluon with momentum $p\gg T$ has
only a finite lifetime,
\beq\label{tbr}
\tbr(p)= \frac{1}{\alpha_s}\sqrt{\frac{p}
 {\hat q}}\,,\eeq
until it disappears via a {\em democratic branching}, i.e. until it splits into a pair of gluons
which carry comparable fractions of the energy $p$ of their parent gluon \cite{Baier:2000sb,Kurkela:2011ti,Blaizot:2013hx}. The emergence of this scale can be understood as follows: the probability
$ \Delta \mcal{P}$ for a branching to occur during an interval $\Delta t$ can be estimated as
\cite{Baier:1996kr,Zakharov:1996fv}
\beq\label{Pdeltat}
 \Delta \mcal{P}\,\simeq\,\alpha_s\frac{\Delta t}{\tf}\,\simeq\,\alpha_s\sqrt{\frac{\hat q}{p}}\,\Delta t\,,
 \eeq
where $\tf\simeq p_z/p_\perp^2$ is the `formation time', i.e. the quantum-mechanical duration 
of a branching process where the softest emitted gluon has 3-momentum $\bp=(\bp_\perp,p_z)$
with $p_z\simeq p\gg p_\perp$. For a branching occurring in the vacuum, $p_z$ and $p_\perp$ are
independent variables, but in the presence of the medium, the emitted gluon acquires
a transverse momentum $p_\perp^2\simeq\hat q\tf$ during the formation time. Using these
two relations, $\tf\simeq p/p_\perp^2$ and $p_\perp^2\simeq\hat q\tf$, one deduces
$\tf(p)\simeq \sqrt{p/\hat q}$, which explains the second equality in \eqref{Pdeltat}. 
This probability $ \Delta \mcal{P}$ become of order one
after a time  $\Delta t\simeq \tbr(p)$, cf. \eqn{tbr}.

Using $\hat q\sim \alpha_s^2 T^3\ln(1/\alpha_s)$, it is easy to check that 
\begin{align}\label{hierarchy}
\tth(p)\,\gg\,\tbr(p)\,\gg\,\trel\qquad \mbox{for}\qquad p\,\gg\,T\,.\end{align}
Hence, a hard gluon with $p_0\gg T$ disappears via democratic branchings   
before having the time to lose a substantial fraction of its energy via drag. In turn,
the daughter gluons will split again and again, thus eventually  producing a gluon cascade. 
Each new gluon generation in this cascade has
a lower energy and hence a shorter lifetime than the previous ones. 
Accordingly, the overall lifetime of the cascade is of the order of the branching time 
$\tbr(p_0)$ of the initial gluon. 

The cascade stops when the branching products become as
soft as the medium constituents: $p\sim T$. Indeed, the soft gluons from the jet
can efficiently thermalize via elastic collisions, over a time interval $\trel$
which is comparable with the corresponding branching time: $\tbr(T)\sim \trel$.
In thermal equilibrium, the detailed balance principle ensures that 
splitting ($1\to 2$) and recombination ($2\to 1$) processes exactly compensate
each other, so the inelastic collision term vanishes, so like the elastic one:
$\mcal{C}_{\rm el}[f]=\mcal{C}_{\rm br}[f]=0$ for $f=f_{\rm eq}$.

Via thermalization, the whole energy $p_0$ of the initial gluon is ultimately transmitted
to the medium.  As anticipated, the characteristic time for thermalization is fixed by 
the branching dynamics and is of order $\tbr(p_0)$. This time is much shorter 
(when $p_0\!\gg\! T$) then the
collisional time scale $\tth(p_0)$  --- the would-be thermalization time  
in the absence of branchings. 

So far, we have implicitly assumed that the lifetime $\tbr(p_0)$ of the cascade
is smaller than the size $L$ of the medium which is available to the jet along the
longitudinal ($z$) axis. Together with \eqref{tbr}, this condition implies an upper limit 
on the momentum $p_0$ of the primary gluon: $p_0\lesssim \omega_{\rm br}(L)\equiv
\alpha_s^2\hat q L^2$. For the conditions at the LHC, it turns out
that this medium scale $\omega_{\rm br}(L)$ is only moderately hard: using typical values 
like $\hat q=1$GeV$^2$/fm, $L=5$\,fm, and $\alpha_s=0.3$, one finds
$\omega_{\rm br}(L)\simeq 12$\,GeV. This is smaller than the energy $E\gtrsim 100$\,GeV
of the LP, but larger than the medium temperature $T\lesssim 1$\,GeV.

The fact that $E\gg \omega_{\rm br}(L)$ implies that the LP cannot disappear inside
the medium: it rather emerges in the final state, 
with a reduced energy though. On the other hand, the LP can abundantly emit
relatively soft  primary gluons, with momenta $T\ll p_0\lesssim \omega_{\rm br}(L)$,
which then generate {\em mini-jets} via democratic branchings. The energy carried by
these mini-jets is eventually transmitted to the medium, as already explained.
The total energy lost by the jet in this way can be estimated as
\cite{Blaizot:2013hx,Fister:2014zxa}
\beq\label{DEtherm}
{\Delta E}_{\rm therm}\,\simeq\,\nu  \, \omega_{\rm br}(L) \,=\,
\nu\, \alpha_s^2 \hat q L^2
\,,\eeq
where  $\nu\simeq 2.5$ is the average number of primary gluons with the
hardest possible energy, i.e. $p_0=\omega_{\rm br}(L)$.

During most stages of the branching process, the cascade is built with relatively 
hard gluons, which are nearly collinear with the LP:
$p_z\gg p_\perp$. It therefore makes sense to focus on the {\em longitudinal} dynamics,
as obtained after integrating out the transverse phase-space. 
This motivates
the following, relatively simple, kinetic equation, for the
{\em longitudinal gluon distribution} $f_\ell(t,z,p_z)\equiv
\int{\rmd^2\bx_\perp \rmd^2\bp_\perp}\,f(t,\bx,\bp)$ \cite{Iancu:2015uja} :
\begin{align}\label{eqL}
&\left(\partial_t+v\partial_z\right)f_\ell(t,z,p)
 = \frac{\hat{q}}{4}{\partial_p}\left[\left(\partial_p + \frac{v}{T}\right) f_\ell(t,z,p) 
 \right]\\*[0.3cm]
&
+\frac{1}{\tbr(p)}\int\limits_r \rmd x\,  {\cal K}(x)\left[\frac{1}{\sqrt{x}}\,
f_\ell\left(t,z,\frac{p}{x}\right) - \frac{1}{2} f_\ell(t,z,p) \right]. \nonumber
\end{align}
Here, $p\equiv p_z$, $v\equiv p/|p|$, $x$ is the splitting fraction, with $0\le x \le 1$,
 ${\cal K}(x)\!\equiv\! {[1-x(1-x)]^{\frac{5}{2}}}/{[x(1-x)]^{\frac{3}{2}}}$
is the BDMPSZ kernel \cite{Baier:1996kr,Zakharov:1996fv},
and the subscript $r$ on the integral over $x$ means that the branching process
is cut off at the soft scale $p=T$.
The two terms within the inelastic collision integral are recognized as the {\em gain}
term and {\em loss} term, respectively.
The initial condition reads
\beq\label{init}
f_\ell(t=0,z,p)=\delta(p-E)\delta(z)\,,
\eeq
corresponding to a LP with longitudinal momentum $p=E$ which enters the medium
at $t=0$ and $z=0$.

\section{The longitudinal gluon distribution}
\label{sec:sol}

\begin{figure}[t]
\begin{center}
 {\includegraphics[width=0.45\textwidth]{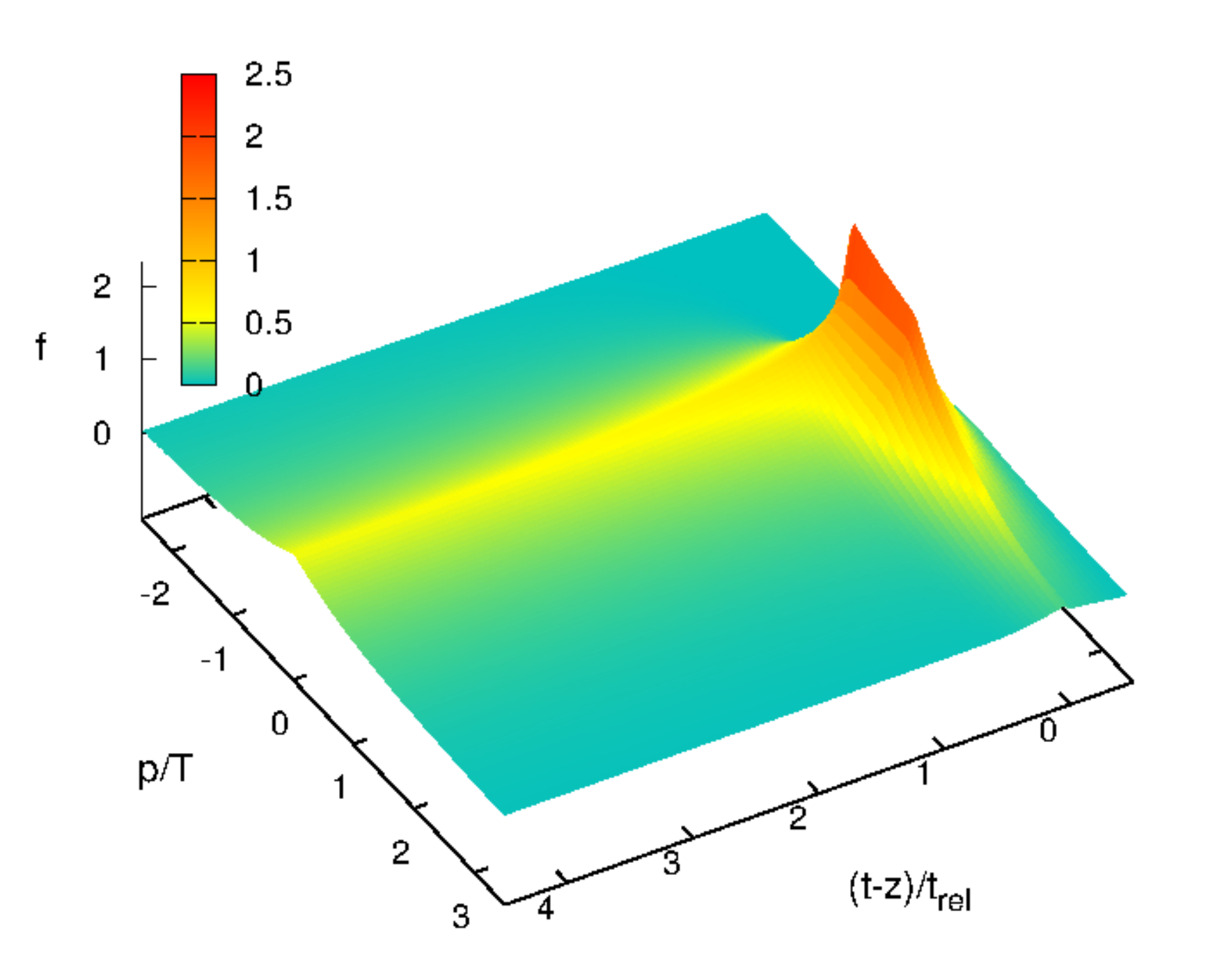}}
\end{center}
\caption{The longitudinal distribution produced by the steady source \eqref{steady} 
shown as a function of $t-z$ and $p$, for $p_0 = T$.}\label{fig:source}
\end{figure}

In this section, we shall present some of the results obtained in \cite{Iancu:2015uja}
via analytic and numerical studies of \eqn{eqL}.
Before we address the general equation \eqref{eqL}, we consider a simpler, related,
problem, where the inelastic collision term in the r.h.s. of \eqref{eqL} 
is replaced by a steady source for particles
with longitudinal momentum $p_0>T$ which propagate at the speed of light:
\begin{align}\label{steady}
\mcal{C}_{\rm br}[f_\ell]\,\to\,T\delta(p-p_0)\delta(z-t)\,.
\end{align}
With $p_0\sim T$, this source mimics the effects of the branching term in so far as the
distribution of the soft gluons ($|p|\lesssim T$) is concerned. This source problem turns out
to be exactly solvable \cite{Iancu:2015uja}, with the result illustrated in Fig.~\ref{fig:source}.
One sees a two component structure, with a front and a tail. The {\em front} is made with
relatively hard gluons, with momenta $T\lesssim p \lesssim p_0$, which propagate
at the speed of light: $z=t$. These are particles injected by the source which have only partially lost
their energy via drag. The {\em tail} lies behind the front ($z< t$) and is made with 
particles which have thermalized under the combined effect of drag and diffusion.
The distribution far behind the front is simply the (one-dimensional) thermal distribution:
\beq\label{MB}
f_\ell(t,z, p) \,\simeq\, \frac{1}{2} \,\rme^{-|p|/T}\,\quad\mbox{when}\ \ t-z\gtrsim \trel\,.
\eeq

\begin{figure}[h]
\begin{center}
\includegraphics[width=.45\textwidth]{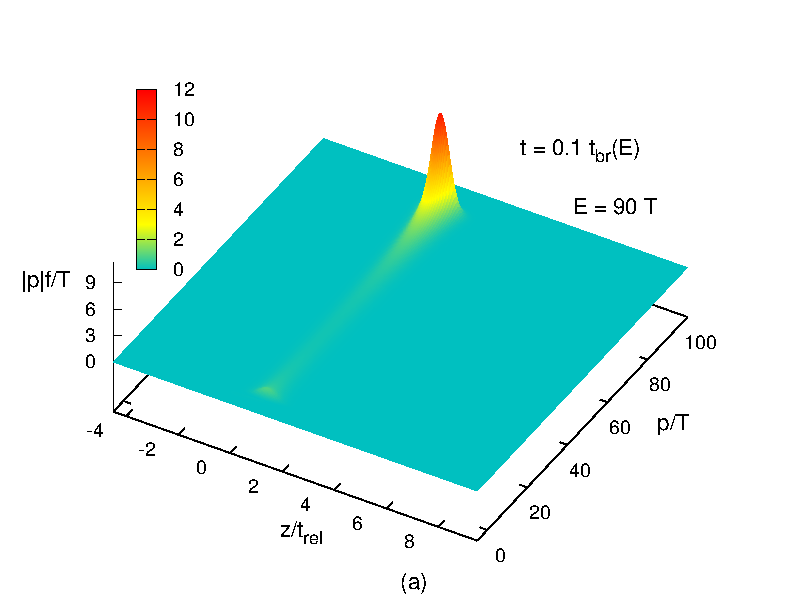}
\includegraphics[width=.45\textwidth]{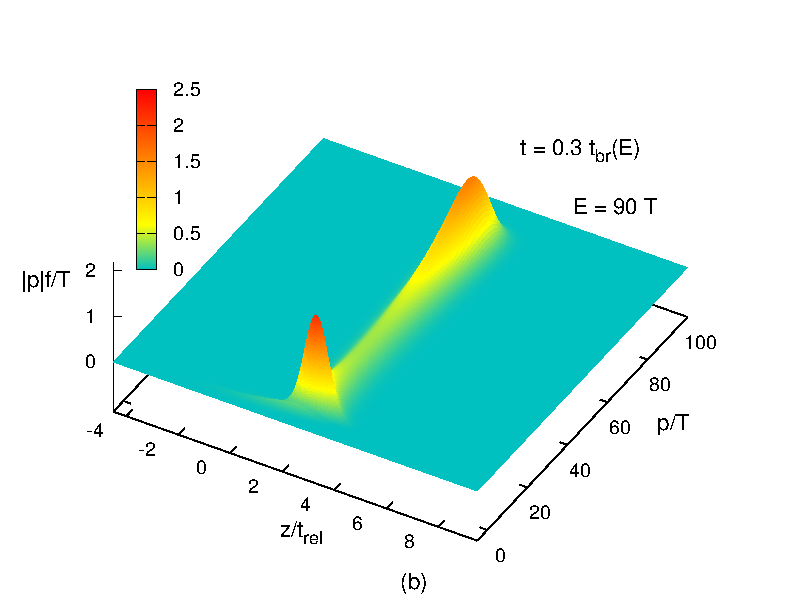}
\end{center}
\caption{The phase-space energy distribution $|p|f_{\ell}/T$ produced by an energetic
jet with $E=90\,T$ at two successive times: (a) an early time $t=0.1\tbr(E)$, when the jet is almost
unquenched; (b) a larger time $t=0.3\tbr(E)$, when the jet is partially quenched.}
\label{fig:early}
\end{figure}

We now turn to the general equation \eqref{eqL} with the initial condition \eqref{init}.
The time scales inherent in this equation are the branching time $\tbr(E)$ for the LP
and the relaxation time $\trel$ via elastic collisions.
In the experimental situation at the LHC, one has $L<\tbr(E)$, as already mentioned,
hence the LP is expected to survive in the final state. This is indeed visible in the numerical results
displayed in Fig.~\ref{fig:early}, as numerically obtained for $E=90\,T$ 
and $\tbr(E)\simeq 10\,\trel$ \cite{Iancu:2015uja}.

Fig.~\ref{fig:early}.a shows the distribution at the very early time $t=0.1\tbr(E)$,
when most of the energy is still carried by the LP. Hence the energy distribution $|p|f_{\ell}/T$
shows a pronounced peak at $p/T=90$ and at $z=t$. Yet, this peak shows some spreading in
$p$, as a consequence of early emissions, which are necessarily soft: the typical quanta emitted
up to time $t$ have $p\lesssim \omega_{\rm br}(t)=
\alpha_s^2\hat q t^2$. 

\begin{figure}[t]
\begin{center}
\includegraphics[width=0.45\textwidth]{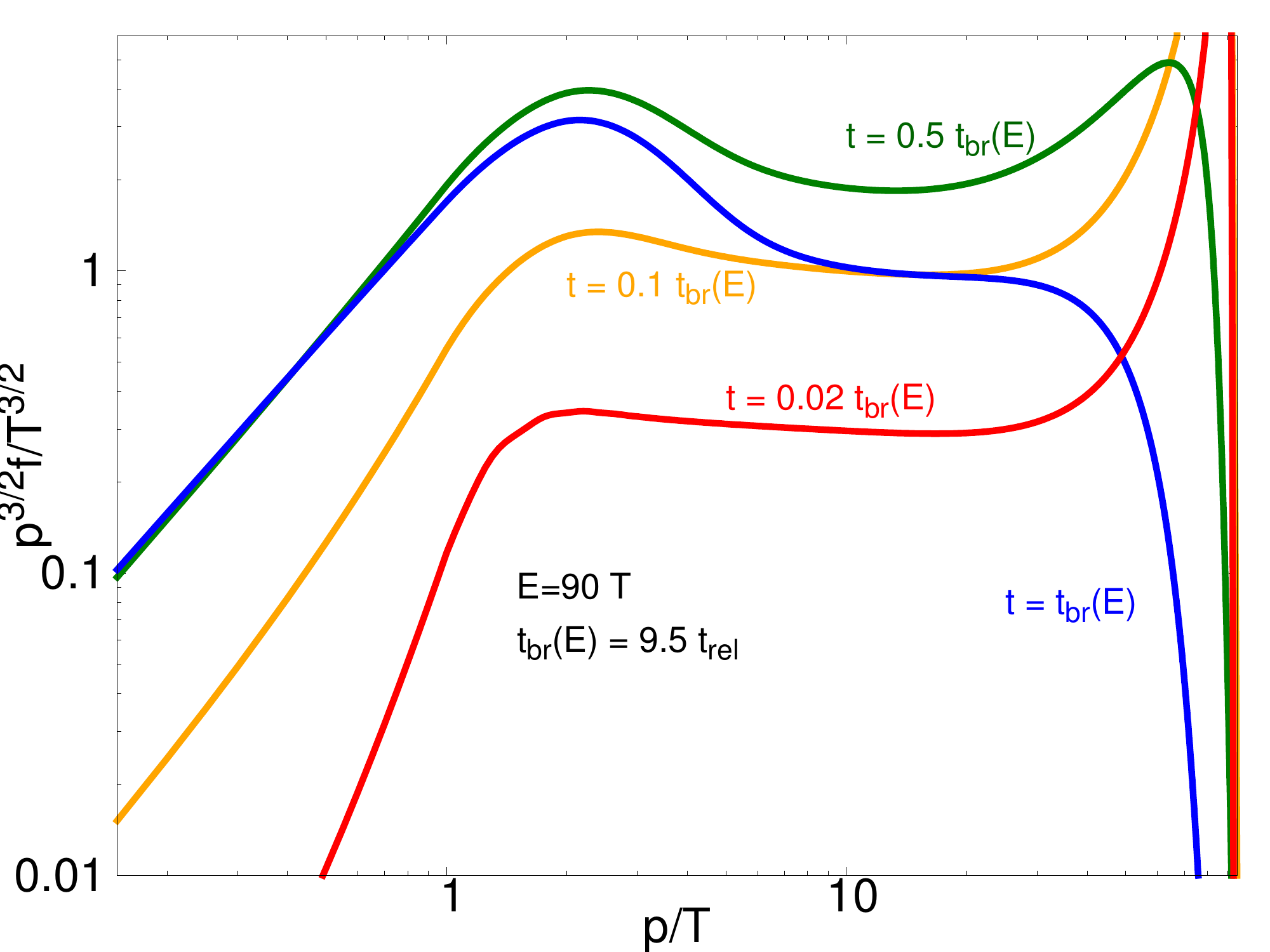}
\caption{The gluon distribution in $p$ at $t=z$ for $E=90\,T$ and for 4 values of time.
The figures show a rather broad window of approximate scaling behavior, $f_{\ell}\propto {1}/{p^{3/2}}$,
at not too large times $t\lesssim \tbr(E)$.
}\label{fig:deviScal}
\end{center}
\end{figure}

At the larger time $t=0.3\tbr(E)$, cf. Fig.~\ref{fig:early}.b, the softening
of the distribution in $p$ is clearly visible, albeit a pronounced LP peak still exists. One can
now distinguish the characteristic `front' + `tail' structure anticipated in Fig.~\ref{fig:source}.
The front at $z=t$ involves relatively hard gluons, whose momentum distribution (within
the range $T < p \ll E$) is given by the scaling spectrum $f_{\ell}\propto
1/p^{3/2}$, as expected for quasi-democratic branchings  
\cite{Baier:2000sb,Blaizot:2013hx}. This scaling behavior, which is a signature
of {\em wave turbulence} \cite{Blaizot:2013hx,Fister:2014zxa}, is better visible in Fig.~\ref{fig:deviScal},
which shows the function $(p/T)^{3/2}f_{\ell}$ at $z=t$.  
Similar results have been obtained in a kinetic theory study of the thermalization 
of the quark-gluon plasma \cite{Kurkela:2014tea}.

The front in Fig.~\ref{fig:early}.b also shows 
a secondary peak at $p=T$, due to the accumulation of gluons
at the lower end of the cascade. Such gluons are abundantly produced via branchings and they
cannot thermalize instantaneously --- rather, they need a time $\sim\trel$ to that aim.
Yet, since $t=0.3\tbr(E)\simeq 3\trel$ is relatively large compared to $\trel$, a thermalized tail
at $z\lesssim t-\trel$ develops indeed, as visible too in Fig.~\ref{fig:early}.b.
This tail carries the energy lost by the jet towards the medium. The numerical studies
demonstrate that this energy loss grows with time like $t^2$, in agreement
with \eqn{DEtherm}  \cite{Iancu:2015uja}.

These results allow for a qualitative comparison with the phenomenology of di-jet asymmetry 
at the LHC  \cite{Aad:2010bu,Chatrchyan:2011sx}. (For a more quantitative comparison,
one could take $T=0.5\,{\rm GeV}$ and 
$\trel=1\,{\rm fm}$.)
 The early situation in Fig.~\ref{fig:early}.a, where the jet is essentially
 unquenched, is illustrative for the leading jet, which crosses at most a very narrow slab of matter. 
 The situation in Fig.~\ref{fig:early}.b, where the jet looks partially quenched,
is representative for the subleading jet in a di-jet event characterized 
by a large asymmetry.  For even larger times, $t\gtrsim \tbr(E)$,
both the LP and the front would disappear and the whole energy would be found in the thermalized tail
  \cite{Iancu:2015uja}.








\end{document}